\documentclass[10pt]{iopart}
\usepackage{graphicx}
\usepackage{epstopdf}
\usepackage{bm}
\usepackage[center]{subfigure}
\usepackage[colorlinks=true, linkcolor=blue]{hyperref}

\usepackage{cite} 

\begin{document}


\title[Mass differential effects in magnetized rotating plasmas for separation applications]{Harnessing mass differential confinement effects in magnetized rotating plasmas to address new separation needs}

\author{R. Gueroult}

\address{Laplace, Universit\'{e} de Toulouse, CNRS, INPT, UPS, 31062 Toulouse, France}
\ead{renaud.gueroult@laplace.univ-tlse.fr}

\author{J.-M. Rax}

\address{Universit\'{e} de Paris XI - Ecole Polytechnique, LOA-ENSTA-CNRS, 91128 Palaiseau, France}

\author{S. J. Zweben, N. J. Fisch}

\address{Princeton Plasma Physics Laboratory, Princeton University, Princeton, NJ 08543 USA}

\begin{abstract}
The ability to separate large volumes of mixed species based on atomic mass appears desirable for a variety of emerging applications with high societal impact. One possibility to meet this objective consists in leveraging mass differential effects in rotating plasmas. Beyond conventional centrifugation, rotating plasmas offer in principle additional ways to separate elements based on mass. Single ion orbits show that ion radial mass separation in a uniform magnetized plasma column can be achieved by applying a tailored electric potential profile across the column, or by driving a rotating magnetic field within the column. Furthermore, magnetic pressure and centrifugal effects can be combined in a non-uniform geometry to separate ions based on mass along the field lines. Practical application of these separation schemes hinges on the ability to produce the desirable electric and magnetic field configuration within the plasma column.

\end{abstract}
\submitto{\PPCF}

\maketitle

\ioptwocol

\section{Introduction}

New and innovative separation techniques could prove extremely valuable in a variety of applications. For instance, it was recently shown that substituting membrane-based separation for distillation in separation processes could lead to a $7\%$ decrease of the total U. S. energy consumption~\cite{Sholl2016}. The same study projects that implementing energy efficient separation techniques in the U. S. petroleum, chemical and paper manufacturing sectors alone could save $100$ million tonnes of carbon dioxide emissions and $\$4$~billion in energy costs annually. Besides the economical incentive, the development of innovative separation techniques is also motivated by their anticipated enabling role in many applications~\cite{NAP1998,ORNL2005}.

Physical separation techniques rely on differences in physical properties. Common examples are distillation, centrifugation and filtration, for which differences in respectively boiling point, mass and size are used. In essence, physical separation techniques harness differential transport and equilibrium properties in a species mixture. A sub-group of physical separation techniques is plasma separation techniques, where the feed to be separated is first turned into a plasma~\cite{Eastland1968}. By ionizing the input feed, separation is carried out at the elemental level. As a result, the whole range of plasma transport phenomenon can in principle be leveraged to produce the desired separation properties.

Physical separation at the elemental level can be traced back to Dempster's mass spectrometer~\cite{Dempster1918} and the calutron device~\cite{Lawrence1958}, in which magnetic deflection was used to separate ions based on mass. In these devices, throughput is limited both by space charge effects~\cite{Smith1947,Parkins2005} and instabilities~\cite{Alexeff1978}. Since plasmas offer a natural neutralization mechanism, plasmas began being considered for separation applications. In particular, the realization that diffusion in a multi-ion species plasma subjected to centrifugal or gravitational forces exhibits asymmetrical effects~\cite{Bonnevier1966,Bonnevier1971,Lehnert1971,ONeil1981} led to the development of plasma centrifuges~\cite{Lehnert1973,Krishnan1981,DelBosco1991} which were then used for isotope separation~\cite{James1976,Krishnan1983,Prasad1987,Hirshfield1989}. In these devices, collisional drag between species leads to an inward drift of the light species and an outward drift of the heavy species~\cite{Bonnevier1966}. Separation arises from the mass dependent, and therefore species dependent, radial equilibrium density profile controlled by rotation~\cite{ONeil1981,Bittencourt1987}. Although plasma centrifuges are conceptually similar to gas and liquid centrifuges, rotation in these devices results from electromagnetic forces and not from frictional entrainment by moving parts. This difference allows for much larger rotation speeds, which in principle translates to higher separation power per centrifuge~\cite{Fetterman2011b}. 

Besides the differential collisional drag exploited in plasma centrifuges, various other differential mechanisms in plasmas were proposed and studied to separate isotopes~\cite{Grossman1991}. For example, differences in excitation energy were used in atomic vapor and molecular laser separation~\cite{Bokhan2006}, while ion-wave interactions, such as ion cyclotron resonance (ICR)~\cite{Dawson1976,Rax2007}, ponderomotive force~\cite{Festeau1985,Eiedler1986} and hybrid resonance~\cite{Hirshfield1976} were suggested for the development of electromagnetic separators. It is worth noting that both the ICR process through the TRW program~\cite{Chen1991a} and the laser separation process through the AVLIS~\cite{Feinberg1993} and MLIS~\cite{Jenson1982} programs were demonstrated in laboratory at large scale.

Isotope separation stands out from other separation needs owing to the small mass difference between the elements to be separated. A legitimate question is therefore to ask whether new plasma mechanisms can be put forward to efficiently separate elements if relaxing the constraint on the mass difference. In this paper, we offer some perspectives on this question. First, in Sec.~\ref{Sec:II}, a variety of applications with high societal impact and for which high-throughput plasma mass separation could prove valuable is highlighted. In light of this observation, in Sec.~\ref{Sec:III}, mass differential confinement properties for a particular class of configurations, namely rotating plasmas, are reviewed. The mass separation potential of rotating plasmas is first considered in a uniform axial magnetic field, and then extended to non-uniform fields. In Sec.~\ref{Sec:IV}, the main findings are summarized.




\section{Rationale for developing high-throughput plasma mass separation techniques}
\label{Sec:II}


\subsection{Need for new separation technologies}

Owing to the role they play in many industries, \emph{e.~g.} chemical, petroleum refining and materials processing, and the opportunities they present for waste reduction and energy efficiency,  the development of separation technologies is of great importance (see, \emph{e.~g.}, Refs.~\cite{NAP1998,ORNL2005}).

For example, consider nuclear waste cleanup in the United States~\cite{Noyes1996}, which is projected to cost more than 280 billion dollars over the next 40 years~\cite{Friedman2014}. Cleanup effort requires notably separating high-activity waste from low-activity waste before vitrification and eventually storage in a geological repository. Due to the heterogeneous and dynamic properties of some of these wastes, conventional separation techniques are particularly challenged. In the meantime, the cleanup cost depends strongly on the efficiency of waste partitioning and minimization~\cite{Swanson1993}.  Therefore, ``\emph{the development of robust and well-understood technologies to enable safe, selective, efficient, and cost-effective cleanup of wastes}" has recently been called for~\cite{OSWEM2016}. 

Another example is nuclear spent fuel reprocessing~\cite{Toumanov2003,Nash2011} as envisioned in advanced nuclear fuel cycles~\cite{ESNIITF2013,OECDNEA2014}. Spent fuel reprocessing and closed fuel cycles, through partitioning and transmuting long lived actinides into shorter lived elements, can decrease the lifetime and associated biological hazards of nuclear spent fuel to a few hundred years~\cite{Magill2003,BlueRibbon2012}. However, transmutation requires separating \emph{a priori} actinides from lanthanides found in spent fuel because of the larger neutron capture cross section of lanthanides~\cite{Mathur2001}. Owing to the similar chemical properties of these elements~\cite{Peterman2008,Morss2011},  separating lanthanides (4-f block elements) from actinides (5-f block elements) requires multiple complex chemical stages, which comes at the expense of cost and reliability~\cite{Gelis2014}. Therefore, a single stage process is desired. 

Finally, yet another example is rare earth recycling. Expansion of rare earth recycling could mitigate the risks associated with the high-volatility of rare earth market~\cite{Bartekova2016}, as well as limit the environmental impact associated with rare earth mining~\cite{Charalampides2016}. However, existing hydro-metallurgical recycling pathways are very similar to the multiple stage processes used for extraction from primary ores~\cite{Binnemans2013}, and hence also often suffer from a significant environmental footprint. Here again, a clean single stage process is desired. 

\subsection{Interest for plasma mass separation}

\begin{figure}
\begin{center}
\includegraphics[]{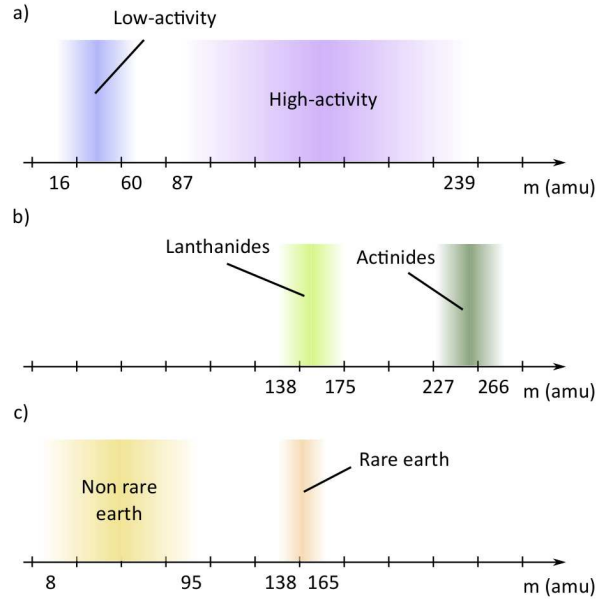}
\caption{Composition of the input feed as a function of atomic mass for various separation needs : a) separation of high activity waste from low activity waste in nuclear waste cleanup, from~\cite{Gueroult2015}, b) actinides/lanthanides separation in nuclear spent fuel reprocessing, from~\cite{Gueroult2014a} and c) rare earth separation in rare earth recycling of NdFeB magnets, from~\cite{Gueroult2017a}. }
\label{Fig:MassSeparation}
\end{center}
\end{figure}

Looking at the elemental composition of the feed and desired product streams of these three separation needs reveals an interesting characteristic.  As shown in Fig.~\ref{Fig:MassSeparation}, the components to be separated break down into a light and a heavy component for each application. For nuclear waste cleanup (Fig.~\ref{Fig:MassSeparation}a), high-activity elements are much heavier than low-activity elements~\cite{Gueroult2015}. For spent fuel reprocessing (Fig.~\ref{Fig:MassSeparation}b), lanthanides are lighter than actinides~\cite{Gueroult2014a}. Finally, for rare earth recovery from NdFeB magnets (Fig.~\ref{Fig:MassSeparation}c), rare earth elements are heavier than all other constituents~\cite{Gueroult2017a}. In all three cases, the light and heavy components are separated by a gap of $30-50$ atomic mass units. In light of this feature, it appears that mass separation at the elemental level could prove valuable for these three applications.

Besides the ability to separate elements adequately, another constrain lies in the ability to process the amount of material required by a given application. For example, take spent fuel reprocessing. Largest chemical reprocessing plants handle some $10^6$~kg of spent fuel per year, of which actinides and lanthanides account for about $10^3$ and $10^4$~kg, respectively~\cite{Bardez2004,Gueroult2014a}. Separation of actinides from lanthanides will thus require a throughput of about $10^4$~kg per year. In plasma mass separation devices designed for isotope separation, a trade-off generally exists between throughput and enrichment factor as a result of collisions. In the TRW experiment~\cite{Chen1991a}, the largest experimental throughput produced was about $10$~kg/year, and the best projections were of the order of $100$~kg/year~\cite[p. 123]{Grossbeck2003}. Although enrichment factor would benefit from larger mass difference allowing in turn for larger throughput, this is unlikely to suffice to meet the requirement of, say, spent fuel reprocessing. Concepts combining ICR selective heating with particle drift in curved magnetic field could possibly achieve $10^3$~kg/year~\cite{Timofeev2014}, but they are still limited by collisions. There is therefore an incentive to look for new plasma mass separation schemes which would allow high-throughput processing.

\subsection{Cost of plasma separation}

Although mass separation at the elemental level holds promise for the applications discussed at the beginning of this section, it remains to show that the cost of plasma separation is not prohibitive. 

After a possible pre-treatment operation, plasma separation first requires turning the input feed into a plasma. In practice, this step breaks down into two sub-steps. The material is first heated to become a gas in the evaporation step. Upon further heating, the gas is turned into a plasma in the ionization step. Once the input feed is ionized, the plasma has to be maintained while separation processes occur. An estimate of the separation cost can be derived based on the cost of each of these processes.

The first process consists in turning the input feed into a plasma.  Strictly speaking, the energy cost for turning solid material into a gas is the sum of the latent heat of both fusion and vaporization, plus the enthalpy change corresponding to heating the material from room temperature to fusion temperature, and then from fusion temperature to boiling temperature. However, the latent heat of vaporization $\mathcal{L}_V$ typically dominates the other contributions, and the energy cost of turning material into a gas is in first approximation $\mathcal{L}_V$. For metals, $\mathcal{L}_V\sim 1-10$~MJ/kg. This however does not include any losses. Using laser ablation as a baseline, the real energy cost for evaporation depends on the laser absorptivity $\chi$, which is typically $0.1-0.4$ for common metals~\cite{Kaplan2014}. Accounting finally for the laser electric efficiency $\eta_l$, the energy cost for evaporation is ${\eta_l}^{-1}\chi^{-1}\mathcal{L}_V$. For a poor $\eta_l=0.1$, this is at most $1$~GJ/kg.  

An estimate for the energy cost of ionization can be obtained by assuming a fully ionized plasma and a given chemical composition. Ionization energy for atoms varies between $3.8$~eV for francium and $24.6$~eV for helium. One kilogram of material of average atomic mass $m_i\sim100$~amu is made of about $6~10^{24}$~atoms. The energy cost for fully ionizing $1$~kg of such material is hence roughly $3-25$~MJ/kg. Here again, energy losses such as excitation and radiation losses need to be accounted for. For helicon plasmas, the efficiency of plasma formation $\eta_p$ has been shown to be about $0.4$ in pure argon~\cite{Lieberman1994}. Since complex plasmas will add extra energy dissipation channels, the plasma efficiency is expected to be lower in this case. For a very degraded $\eta_p = 0.02$, the cost of plasma formation and and maintenance is about $0.15-1.25$~GJ/kg.

Summing these two contributions, an upper bound energy cost for plasma separation is $2$~GJ/kg. For an electricity cost of $\$0.12$ per kW.h, this is about $\$65$ per kg. It is worth pointing out here that this figure could be significantly lower if one could produce separation in a partially ionized plasma since the ionization cost scales with the number of ionized atoms. Besides processing costs considered thus far, capital, operation and maintenance costs will have to be accounted for. However, owing to a comparatively small footprint, capital costs of plasma techniques are  expected to be inferior to those of chemical techniques. This scaling was for example observed when comparing high-temperature processing and aqueous processing~\cite{NationalResearchCouncil2001}.

A processing cost of $\$65$ per kg already suggests that plasma techniques, at least in their current form, are unlikely to be attractive for applications for which proven techniques are readily available.  However, preliminary cost comparative suggests that plasma techniques might be competitive with the proposed chemical solutions for nuclear waste cleanup thanks to improved waste minimization~\cite{Gueroult2015}. Similar considerations suggest that plasma processing of NdFeB magnets for rare earth recovery could be economically attractive~\cite{Gueroult2017a}. Finally, the continuous plasma processing of spent fuel unloaded from a nuclear reactor has been estimated to only require $0.06\%$ of the output power of this reactor~\cite{Zhiltsov2006}.  

Further to this point, a complete cost comparison should include the environmental, social and economical costs and benefits of each separation techniques. Although these effects are difficult to quantify, this is where plasma techniques could prove particularly advantageous. It may be that the very limited environmental cost of plasma techniques will offset a possible disadvantage when considering processing costs alone.


\section{Mass differential confinement effects in magnetized rotating plasmas}
\label{Sec:III}

In the search for mass differential effects, rotating configurations hold particular promise thanks to the centrifugal effects associated with rotation. In the remaining of this paper, we therefore analyze in a systematic manner mass differential confinement properties in magnetized rotating plasmas.

\subsection{$E\times B$ rotating plasmas in purely axial magnetic field}

\begin{figure}
\begin{center}
\includegraphics[]{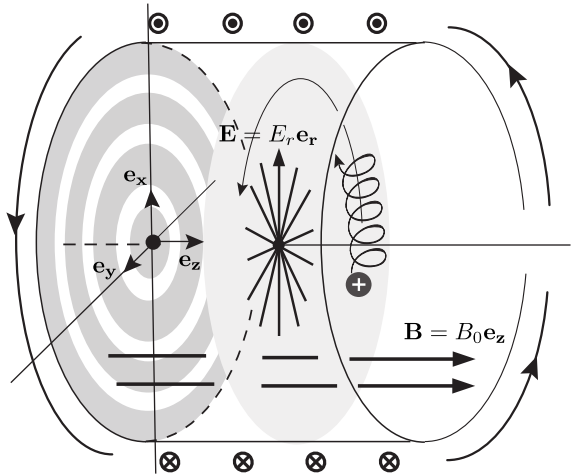}
\caption{Linear configuration: uniform axial magnetic field and radial electric field. }
\label{Fig:Linear}
\end{center}
\end{figure}

Two relatively simple configurations can be readily identified to produce rotation as a result of the $\mathbf{E}\times\mathbf{B}$ drift in a plasma column: an axial magnetic combined with a radial electric field, or, alternatively, a radial magnetic field and an axial electric field. The former of these two configurations, with a uniform axial magnetic field $\mathbf{B} = B_0 \mathbf{e_z}$ and a radial electric field $\mathbf{E} = E_r \mathbf{e_r} = -\bm{\nabla}(\phi)$ as depicted in Fig.~\ref{Fig:Linear}, displays the interesting property that the cross-product of the centrifugal force with the magnetic field is non zero. As a result, centrifugal forces cause an additional azimuthal drift. 

\begin{figure}
\begin{center}
\includegraphics[]{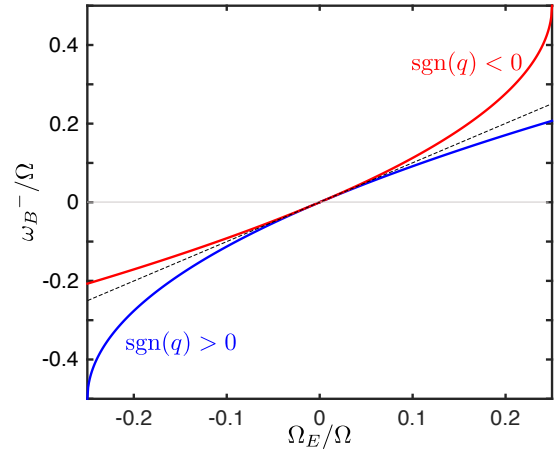}
\caption{Slow Brillouin mode for positively (blue) and negatively (red) charged particles, with $\Omega = |q|B_0/m$ the gyro-frequency and $\Omega_E = -E_r/B_0$ the $\mathbf{E}\times\mathbf{B}$ drift angular frequency. Rotation is counter-clockwise (${\omega_B}^->0$) for $E_r<0$, and reciprocally. The black dotted curve represents the zero inertia solution. }
\label{Fig:Brillouin}
\end{center}
\end{figure}

Considering the plasma column in Fig.~\ref{Fig:Linear}, and neglecting first collisions, the radial force balance on a particle of charge $q$ and mass $m$ writes
\begin{equation}
-\omega^2 = \frac{q}{mr}E_r+\textrm{sgn}(q)\Omega\omega,
\end{equation}
with $\omega$ the particle azimuthal angular frequency, $\textrm{sgn}(x) = x/|x|$ the sign function and $\Omega = |q|B_0/m$ the cyclotron frequency.
The equilibrium solution is described by the slow and fast Brillouin modes~\cite{Davidson2001}
\begin{equation}
{\omega_B}^{\pm} = -\textrm{sgn}(q)\frac{\Omega}{2}\left[1\pm\sqrt{1-\frac{4mE_r}{q{B_0}^2r}}\right].
\label{Eq:Brillouin}
\end{equation}
Out of these two modes, only the slow mode ${\omega_B}^-$ depicted in Fig.~\ref{Fig:Brillouin} arises spontaneously. Introducing the azimuthal drift velocity in the limit of zero inertia $\Omega_E = -E_r/(rB_0)$, Eq.~(\ref{Eq:Brillouin}) rewrites
\begin{equation}
{\omega_B}^{-} = -\textrm{sgn}(q)\frac{\Omega}{2}\left[1-\sqrt{1+4\textrm{sgn}(q)\frac{\Omega_E}{\Omega}}\right].
\label{Eq:Brillouin2}
\end{equation}
Taylor expanding Eq.~(\ref{Eq:Brillouin2}) for $|\Omega_E|/\Omega\ll 1$, one gets
\begin{equation}
{\omega_B}^- = \Omega_E\left[1-\textrm{sgn}(q)\frac{\Omega_E}{\Omega}+\mathcal{O}\left(\left[\frac{\Omega_E}{\Omega}\right]^2\right)\right]
\label{Eq:TaylorBrillouin} 
\end{equation}
In the limit $|\Omega_E|/\Omega\rightarrow 0$, one recovers $\omega = \Omega_E$. In this limit, there is no difference in azimuthal $\mathbf{E}\times\mathbf{B}$ drift velocity between charged species. 

For $E_r\leq 0$, $\Omega_E\geq 0$, and plasma rotation is in the counter-clockwise direction. Eq.~(\ref{Eq:Brillouin2}) shows that centrifugal effects speed up rotation for negatively charged particles, and slow up particles for positively charged particles, as seen in Fig.~\ref{Fig:Brillouin}. For two positive ions of different mass, the angular velocity of the light ion is larger than the angular velocity of the heavy ion. On the other hand, rotation is in the clockwise direction for $E_r\geq 0$, and $\Omega_E\leq 0$. For this polarity, positively charged particles rotate faster, while negatively charged particles rotate slower. For two positive ions of different mass, the norm of the angular velocity of the light ion is smaller than the norm of the angular velocity of the heavy ion. Note that the difference in azimuthal velocity between two different ion species leads to a positive (resp. negative) azimuthal drag force on heavy (resp. light) ions no matter the polarity of the radial electric field. In both cases, this drag force causes light ions to drift radially inward and heavy ions to drift radially outward. This is the physical mechanism behind plasma centrifugation~\cite{Bonnevier1971}.

Now, looking at Eq.~(\ref{Eq:Brillouin2}), one notices that there is no solution if
\begin{equation}
\textrm{sgn}(q)\frac{\Omega_E}{\Omega}\geq -1/4.
\label{Eq:Brillouin_limit}
\end{equation}
This limit, known as the Brillouin limit, means that ions are radially unconfined for fast enough rotation in the clockwise direction, and electrons are unconfined for fast enough rotation in the counter-clockwise direction. Since $\Omega_e\gg\Omega_i$, the latter is however unlikely. 

To illustrate these confinement properties, it is interesting (as it will become clear later) to consider particle equilibrium in the frame rotating with the angular velocity $\bm{\varpi} = -\textrm{sgn}(q)\Omega/2~\mathbf{e_z}$. Let us denote variables in this rotating frame with a $\tilde{~}$. Since $\partial \varpi/\partial t = 0$, the fields transformation reads
\numparts
\begin{equation}
\tilde{\mathbf{E}} = \mathbf{E} + (\bm{\varpi}\times\tilde{r})\times\mathbf{B}
\end{equation}
\begin{equation}
\tilde{\mathbf{B}} = \mathbf{B}.
\end{equation}
\endnumparts
Assuming the fields in the rotating frame do not depend on time, one can rewrite the Newton-Lorentz equation as (see for example, Ref~\cite[p. 328]{Hestenes1998})
\begin{equation}
m\frac{\partial{\tilde{\mathbf{v}}}}{\partial t} = q({\mathbf{E}}^{\star}+\tilde{\mathbf{v}}\times\mathbf{B}^{\star})
\end{equation}
with
\numparts
\begin{equation}
\mathbf{E}^{\star} = \tilde{\mathbf{E}}+\bm{\nabla}\left(\frac{m\varpi^2\tilde{r}^2}{2q}\right)
\label{Eq:Estar}
\end{equation}
\begin{equation}
\mathbf{B}^{\star} = \tilde{\mathbf{B}}+\frac{2m}{q}\bm{\varpi} = 0.
\label{Eq:B_transform}
\end{equation}
\endnumparts
Eq.~(\ref{Eq:B_transform}) shows that in the chosen frame rotating with the angular frequency $\bm{\varpi} = -\textrm{sgn}(q)\Omega/2~\mathbf{e_z}$, the magnetic field cancels. In this frame, the particle dynamics is only controlled by the electric field $\mathbf{E}^{\star}$. The second term on the right hand side in Eq.~(\ref{Eq:Estar}) is the contribution of the centrifugal force. Introducing  the effective potential
\begin{equation}
\phi^{\star}(\tilde{r}) = \phi(r)+[2\textrm{sgn}(q)-1]\frac{q{B_0}^2}{8m}\tilde{r}^2,
\label{Eq:radialpotentialall}
\end{equation}
Eq.~(\ref{Eq:Estar}) writes $\mathbf{E}^{\star} = -\bm{\nabla}\phi^{\star}$. Since the Coriolis force is proportional to $\varpi$, it depends on the sign of the rotation and therefore here on the sign of the particle charge. In contrast, the centrifugal force is proportional to $\varpi^2$, and is therefore positive irrespective of the sign of the particle charge. For positively charged particles, Coriolis and centrifugal forces are in opposite direction, and one gets
\begin{equation}
\phi^{\star}(\tilde{r}) = \phi(r)+\frac{q{B_0}^2}{8m}\tilde{r}^2.
\label{Eq:radialpotential}
\end{equation}

\begin{figure}
\begin{center}
\subfigure[Laboratory potential $\phi(r)$]{\includegraphics[]{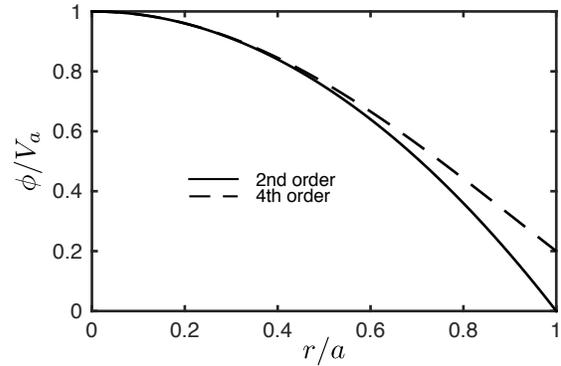}\label{Fig:FieldsA}}\hspace{0.2cm}\subfigure[Effective potential $\phi^{\star}(r)$]{\includegraphics[]{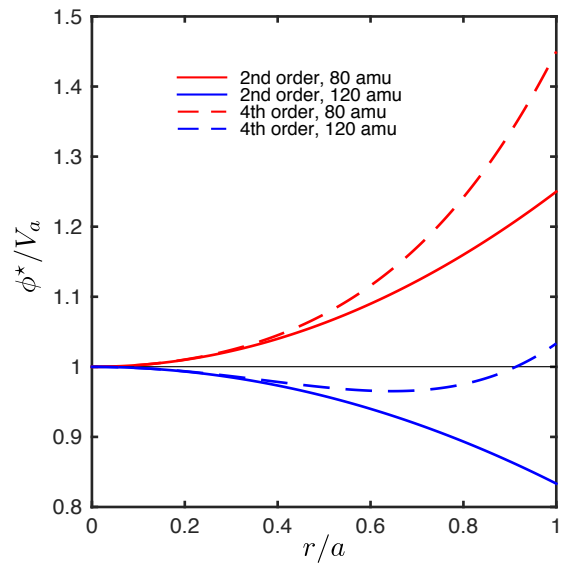}\label{Fig:FieldsB}}
\caption{Applied laboratory potential radial profile $\phi(r)$ [\subref{Fig:FieldsA}] and effective potential radial profile $\phi^{\star}(r)$ for two different ion mass [\subref{Fig:FieldsB}]. Solid line curves are obtained for $\phi(r) = V_a(1- r^2/a^2)$, while dotted line curves are obtained for $\phi(r) = \beta r^4 + V_a(1- r^2/a^2)$, $\beta \in \textrm{I}\!\textrm{R}$. $V_a$ is the potential difference across the plasma column for the parabolic case, $a$ is the plasma column radius. The addition of a fourth order term to the parabolic profile leads to the formation of a potential well off-axis for heavy ions (dotted blue curve). }
\label{Fig:Fields}
\end{center}
\end{figure}

If the potential applied in the laboratory frame is constant ($\partial\phi/\partial r = 0$), the effective potential $\phi^{\star}$ in Eq.~(\ref{Eq:radialpotential}) is convex, and ions are confined. Eq.~(\ref{Eq:radialpotentialall}) shows electrons are also confined in this case. Now assume a parabolic potential profile $\phi(r) = \alpha r^2$ is applied in the laboratory frame. This corresponds to a solid body rotating plasma column since $E_r\propto r$ so $\partial\Omega_e/\partial r = 0$, and thus, using Eq.~(\ref{Eq:Brillouin2}), $\partial{\omega_B}^-/\partial r = 0$. For $\alpha\geq-q{B_0}^2/(8m)$, an ion of mass $m$ and charge $q$ is still confined. On the other hand, for $\alpha\leq-q{B_0}^2/(8m)$, Eq.~(\ref{Eq:radialpotential}) tells us that $\phi^{\star}$ is concave. An ion of mass $m$ and charge $q$ is therefore radially unconfined. The change in concavity of the effective potential profile $\phi^{\star}(r)$ is illustrated in Fig.~\ref{Fig:Fields}. The threshold value $\alpha_c=-q{B_0}^2/(8m)$ for ion confinement can be rewritten $E_r/(r B_0) = \Omega/4$, which is the Brillouin limit given in Eq.~(\ref{Eq:Brillouin_limit}). Now suppose a multi-ion species plasma with $\alpha=-q{B_0}^2/(8m_{\diamond})$, so that
\begin{equation}
\phi^{\star}(\tilde{r}) =\frac{q{B_0}^2}{8m m_{\diamond}}(m_{\diamond}-m)\tilde{r}^2.
\end{equation}
The effective potential $\phi^{\star}$ indicates that a singly charged ion with mass $m\geq m_{\diamond}$ will be radially unconfined, while a singly charged ion with mass $m\leq m_{\diamond}$ will be radially confined. This charge to mass ratio threshold for confinement is the basis for the DC band gap ion mass filter~\cite{Ohkawa2002} used in the Archimedes plasma mass filter~\cite{Freeman2003}. In this device, ions are separated into two components: light ions $m/m_{\diamond}<1$ are collected axially along the magnetic field lines while heavy ions $m/m_{\diamond}>1$ are collected radially.

Practically, this filtering mechanism has a few limitations. First, since the confinement criteria depends on the charge to mass ratio and not on the mass alone: a doubly charged ion of mass  $2m$ can not be differentiated from a singly charged ion of mass $m$. This means that heavy doubly charged ions will be collected with light singly charged ions. Second, the filtering mechanism relies on low collisionality, which sets a limit on plasma density and hence throughput for a given magnetic field intensity. Indeed, the radial ion transport induced by collisions with neutrals brings light ions ($m/m_{\diamond}<1$) to the heavy ions ($m/m_{\diamond}>1$) stream. Strictly speaking, ion-neutral collisions slow down the slow mode and suppress the requirement for ion radial confinement $\Omega_E/\Omega\geq -1/4$~\cite{Rax2015}. In other words, the Brillouin limit breaks down. Separation then hinges on the differential radial transport properties of light and heavy ions. Finally, and maybe most importantly, another limitation is that heavy ions are typically collected over a large region of the plasma chamber. This is because heavy ions are extracted perpendicularly to the field lines with little control. This is particularly an issue when heavy particles are made of hazardous materials, for example for nuclear waste cleanup.

To remediate to this last issue, one can use a higher order polynomial profile for the laboratory potential $\phi$. For example, a fourth order polynomial can be used to create a dip in effective potential off-axis, while maintaining global radial confinement~\cite{Gueroult2014}. This scheme is illustrated in Fig.~\ref{Fig:Fields}. For ions lighter than the mass threshold (red dotted curve in Fig.~\ref{Fig:FieldsA}), the effective potential $\phi^{\star}$ is monotonically increasing with $r$, so that light ions are collected axially along the field lines in the central region. On the other hand, heavy ions see a minimum in effective potential off-axis (blue dotted curve in Fig.~\ref{Fig:FieldsB}), but are still radially confined as opposed to the parabolic potential case. Heavy ions are therefore collected axially along the field lines but in an annular region at larger radius. In contrast with the DC band gap ion mass filter~\cite{Ohkawa2002} for which collisions are detrimental to separation performances, collisions are in this configuration required since they allow radial diffusion of heavy ions towards the off-axis potential well. Without collisions, heavy ions would extend radially from the center to a point $r_h$ past the minimum of $\phi^{\star}$ and which depends on the ion temperature. Radial separation of heavy from light ions can be optimized through the radial profile of the laboratory potential $\phi(r)$ and the device geometry~\cite{Gueroult2014}. However, the advantage of extracting both light and heavy species along field lines comes at the expense of producing and controlling a more complex potential radial profile in the plasma. Also, the use of higher-order potential profile means that plasma rotation is now sheared ($\partial \omega/\partial r\neq 0$), which is known to lead, under certain conditions, to the onset of Kelvin-Helmholtz instabilities~\cite{Kent1969,Jassby1972}.

To conclude this discussion of mass separation due to $\mathbf{E}\times \mathbf{B}$ rotation in a uniform magnetic field, it is worth noting that both radial-axial mass separation in Archimedes filter~\cite{Freeman2003} and radial-radial mass separation in the double well mass filter~\cite{Gueroult2014} require fast plasma rotation. Quantitatively, separation occurs near the Brillouin limit for which $|\omega|/\Omega\sim1/2$. For this rotation regime, significant differences in azimuthal velocities can exist between ions with different mass, which could trigger the onset of centrifugal instabilities~\cite{Rosenbluth1962,Chen1966,Gueroult2017b}. In light of this observation, it seems advantageous if possible to produce mass separation at lower rotation velocity. This might be done by abandoning the uniform axial magnetic field topology considered up to this point.

\subsection{$E\times B$ rotating plasmas in inclined magnetic fields}

Substituting an inclined magnetic field $\mathbf{B} = B_r \mathbf{e_r} + B_z\mathbf{e_z} = B \mathbf{e_b} $ in place of the purely axial field considered in the previous section offers additional means of control. Assume a conical magnetic surface defined by $\mathbf{B}\cdot\mathbf{e_z}=\cos{\alpha}$, and write $s$ the curvilinear coordinate along a given field line. Flux conservation requires $B(s) = B_0 r_0/r(s)$, where $B_0$ and $r_0$ are the magnetic field intensity and the field line radius at $s=0$, respectively. In this configuration, the magnetic mirror force
\begin{equation}
-\mu\frac{\partial B}{\partial s}\mathbf{e_b} = \mu B_0\frac{r_0}{r(s)^2}\sin\alpha\mathbf{e_b},
\label{Eq:mirror_force}
\end{equation}
with $\mu = m{v_{\perp}}^2/(2B)$ the magnetic moment of the particle. Besides this mirror force, the force balance along the field line requires accounting for the contribution of centrifugal forces produced by plasma rotation  $m\omega^2r(s)\sin\alpha\mathbf{e_b}$, with $\omega$ the rotation velocity. Note that the iso-rotation theorem states that $\omega$ is constant on a given magnetic surface (see, \emph{e.~g.}, Refs.~\cite{Ferraro1937,Lehnert1971}). Centrifugal and mirror forces add up, and a particle moving along the field line towards larger $r$ accelerates, while a particle moving towards smaller $r$ slows down.

\begin{figure}
\begin{center}
\subfigure[Centrifugal end plug created by an inclined magnetic field]{\includegraphics[]{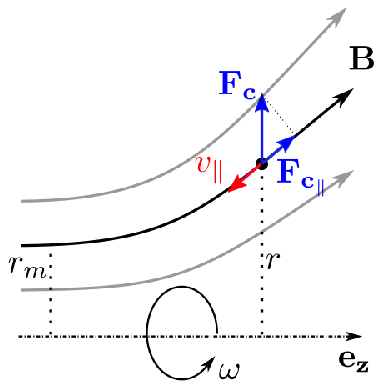}\label{Fig:Diverging}}\subfigure[Collection diagram at $r_m$ for heavy and light ions in thermal equilibrium]{\includegraphics[]{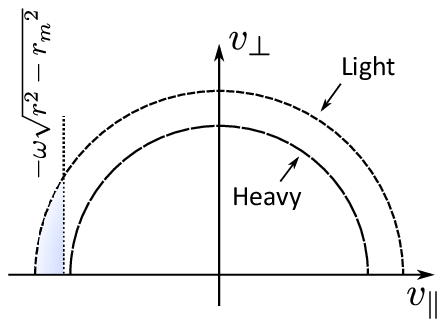}\label{Fig:DivergingLoss}}
\caption{Magnetic field topology [\subref{Fig:Diverging}] and mass separation capabilities~[\subref{Fig:DivergingLoss}] of a centrifugal end plug. The blue shaded region in Fig.~\subref{Fig:DivergingLoss} represents the part of ions starting at a radius $r$ which are collected at the minimum radius $r_m$ along the same field line. Mirror forces are here neglected. }
\label{Fig:NonLinear1}
\end{center}
\end{figure}

Consider now the field topology depicted in Fig.~\ref{Fig:Diverging}. A particle at radius $r$ with negative $v_{\parallel}$ sees a centrifugal potential barrier $m\omega^2(r^2-{r_m}^2)/2$. Interestingly, this potential barrier is proportional to the particle mass. For a given parallel energy $\epsilon_{\parallel}$ and two particles of mass $m_l$ and $m_h$ with $m_l\leq m_h$, there therefore exists a rotation velocity $\omega$ for which the light particle can reach $r_m$, while the heavy particle can not. Assuming a two-ion species plasma in thermal equilibrium, this result can in principle be used to preferentially collect light ions on the left side in Fig.~\ref{Fig:Diverging}, as illustrated in Fig.~\ref{Fig:DivergingLoss}. Strictly speaking, one should also consider the mirror force which tends to pull particles to larger $r$. However, from Eq.(\ref{Eq:mirror_force}), the ratio of centrifugal to mirror forces along the field line is
\begin{equation}
\frac{\mathbf{F_c}\cdot\mathbf{e_b}}{-\mu\partial B/\partial s} = \frac{2r^3\omega^2}{r_0{v_{\perp_0}}^2},
\end{equation}
and mirror effects should be negligible for large enough $r$.

\begin{figure}
\begin{center}
\subfigure[Mirror end plug at large radius ]{\includegraphics[]{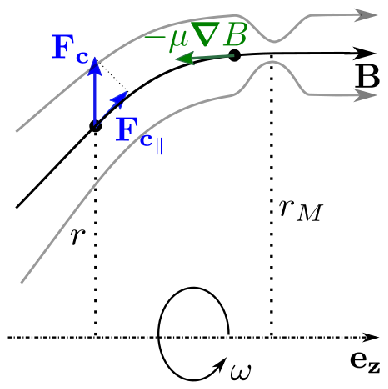}\label{Fig:Mirror}}\subfigure[Particle confinement for heavy and light ions in thermal equilibrium]{\includegraphics[]{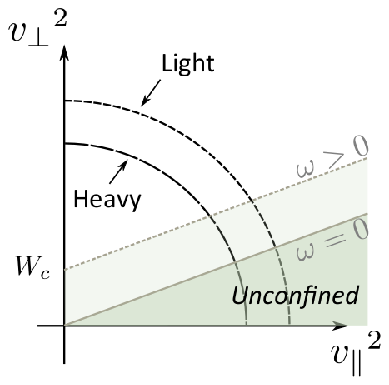}\label{Fig:MirrorLoss}}
\caption{Magnetic field topology [\subref{Fig:Mirror}] and mass separation capabilities [\subref{Fig:MirrorLoss}] of a mirror end plug. $W_c$  is defined in Eq.~(\ref{Eq:threshold_perp_velocity}). The shaded area in Fig.~\subref{Fig:MirrorLoss} represents the mirror loss cone, which grows with the rotation velocity $\omega$. }
\label{Fig:NonLinear2}
\end{center}
\end{figure}

Centrifugal and mirror effects can also be used in a way that they oppose each other, for example by creating a magnetic mirror at a larger radius as illustrated in Fig.~\ref{Fig:Mirror}. In this geometry, energy conservation yields the confinement criteria~\cite{Lehnert1971,Volosov2006}
\begin{equation}
{v_{\parallel}}^2\leq (r\Omega)^2\left(1-\frac{{r_M}^2}{{r}^2}\right)+{v_{\perp}}^2\left(\frac{B_M}{B}-1\right),
\label{Eq:mirror_confinement}
\end{equation}
where $r_M$ and $B_M$ are the field line radius and field intensity at the mirror, respectively. In the configuration depicted in Fig.~\ref{Fig:Mirror}, $r_M/r>1$, so that the first term on the right hand side in Eq.~(\ref{Eq:mirror_confinement}) is negative, while the second term on the right hand side is positive. As a result, a particle with $v_{\parallel}=0$ is only confined if ${v_{\perp}}^2\geq W_c$, with
\begin{equation}
W_c = \left[\left(\frac{{r_M}}{{r}}\right)^2-1\right]\left(\frac{B_M}{B}-1\right)^{-1}{r}^2\omega^2.
\label{Eq:threshold_perp_velocity}
\end{equation}
This is in contrast with conventional rotating mirror machines ($r_M/r\ll1$) in which centrifugal forces combine with mirror forces to enhance ion confinement~\cite{Lehnert1974,Bekhtenev1980}. The threshold in perpendicular velocity described by Eq.~(\ref{Eq:threshold_perp_velocity}) creates mass differential confinement properties as shown in Fig.~\ref{Fig:MirrorLoss}. Considering again a two-ion species plasma in thermal equilibrium, the loss cone modified by rotation will cover a larger fraction of the distribution of heavy ions compared to the distribution of light ions. The rotation velocity $\omega$ provides control over the fraction of heavy ions lost through the mirror.

These two effects, namely preferential collection of light ions at smaller radius (Fig.~\ref{Fig:NonLinear1}) and preferential collection of heavy ions through a magnetic mirror at large radius (Fig.~\ref{Fig:NonLinear2}), are the basis of the Magnetic Centrifugal Mass Filter (MCMF)~\cite{Fetterman2011}. In this device, collisionality has to be large enough for ion-ion pitch angle scattering to scatter ions into the small radius side loss cone, but low enough to limit perpendicular transport. 
The mass separation capabilities were confirmed through preliminary numerical simulations~\cite{Gueroult2012a,Gueroult2014a}, and constrains imposed by collisionality on the operating window were recently highlighted~\cite{Ochs2017}. 

One critical question on which hinges the demonstration of the practicality of $\mathbf{E}\times\mathbf{B}$ rotating plasma configurations for mass separation is the ability to establish and control the required perpendicular electric field in the plasma via end electrodes (see, \emph{e.~g.}, Refs~\cite{Tsushima1986,Shinohara2007,Gueroult2016a}). Alternatively, wave-induced rotation has been suggested~\cite{Fetterman2009} as a way to suppress the need for end electrodes, but remains to be validated experimentally. Another possibility to produce plasma rotation might lie in the use of rotating magnetic fields.

\subsection{Plasma rotation in rotating magnetic fields}

\begin{figure}
\begin{center}
\includegraphics[]{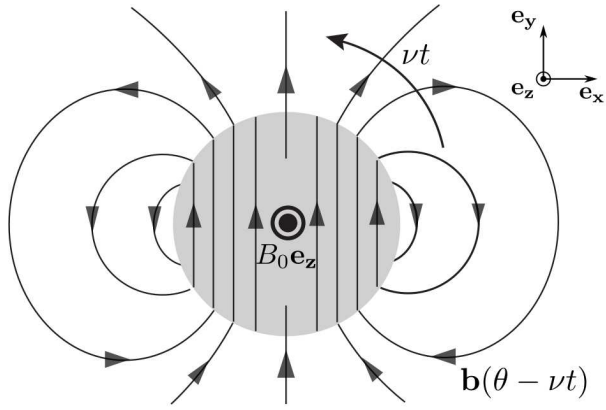}
\caption{Linear configuration: static axial magnetic field $B_0\mathbf{e_z}$ plus rotating magnetic field $\mathbf{b} = b_0 [\cos(\nu t)\mathbf{e_x}+\sin(\nu t)\mathbf{e_y}]$. }
\label{Fig:Rotating}
\end{center}
\end{figure}

Alfv{\'e}n's frozen in theorem predicts that a magnetized plasma column with an axial static magnetic field $B_0\mathbf{e_z}$ can, under certain conditions, be spun using a rotating magnetic field~\cite{Stepanov1958}. This configuration is depicted in Fig.~\ref{Fig:Rotating}. However, single particle dynamics in rotating field configurations is far more convoluted than simple rotation~\cite{Soldatenkov1966,FIsch1982,Hugrass1983}. It was recently shown that ion confinement in rotating depends strongly on the particular external driving currents geometry used to produce this rotating magnetic field~\cite{Rax2016}. This result can be understood by noting that the rotating magnetic field 
\begin{equation}
\mathbf{b} = b_0 [\cos(\nu t)\mathbf{e_x}+\sin(\nu t)\mathbf{e_y}]
\end{equation}
can be obtained from any linear combination of the two vector potentials 
\numparts
\begin{equation}
\mathbf{A_s} = b_0 \left[y\cos(\nu t)-x\sin(\nu t)\right]\mathbf{e_z},
\end{equation}
\begin{equation}
\mathbf{A_a} = b_0 z \left[\sin(\nu t)\mathbf{e_x}-\cos(\nu t)\mathbf{e_y}\right],
\end{equation}
\endnumparts
and that each of this vector potential combination leads to a different particle dynamic as a result of a different inductive electric field $\mathbf{E} = -\partial \mathbf{A}/\partial t$. Furthermore, the orbit of a particle in a given combination of $\mathbf{A_s}$ and $\mathbf{A_a}$ can not be simply deduced from the particle orbits in $\mathbf{A_s}$ and $\mathbf{A_a}$ separately~\cite{Rax2016}. 

\begin{figure}
\begin{center}
\includegraphics[]{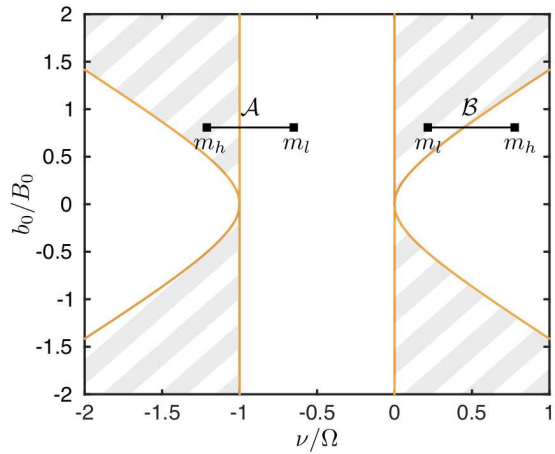}
\caption{Stability diagram for the vector potential $\mathbf{A_s}$, from~\cite{Rax2016}. $\nu$ is the rotating field angular frequency,  $\Omega = |q|B_0/m$ is the gyro-frequency and $B_0$ and $b_0$ are the axial and rotating magnetic field amplitudes, respectively. Hatched regions denote unstable regions, where ions are radially unconfined. $m_l$ and $m_h$ denote the mass of two ions ($m_l < m_h$). Since $\Omega \propto m^{-1}$, there exists $\nu$ such that the light ion is confined while the heavy ion is unconfined (case $\mathcal{B}$), and reciprocally (case $\mathcal{A}$). }
\label{Fig:RotatingStability}
\end{center}
\end{figure}

For each vector potential field, ion stability criteria depends on $\nu/\Omega$ and $b_0/B_0$, with $\Omega = |q|B_0/m$. Since $\Omega\propto m^{-1}$, any stability frontier which is not purely horizontal in the ($\nu/\Omega,b_0/B_0$) plane offers opportunities for mass separation.  For example, the stability diagram for the simple case $\mathbf{A_s}$ is plotted in Fig.~\ref{Fig:RotatingStability}. For any rotating field amplitude $b_0\neq 0$, three regions can be used to separate elements based on mass. For case $\mathcal{B}$ in Fig.~\ref{Fig:RotatingStability}, a heavy ion of mass $m_h$ is radially confined, but for the same conditions a light ion of mass $m_l$ is not. A similar situation is found for the same parameters if driving the rotating field in the opposite direction. On the other hand, light ions are confined while heavy ions are not for case $\mathcal{A}$. The possibility to choose which of heavy or light particles is radially confined could prove very useful for some applications, and addresses one of the limitations of the Archimedes filter~\cite{Freeman2003} discussed earlier in this section. 
 
In light of these results, rotating magnetic field configurations appear promising for mass separation applications. However, it should be stressed that the vector potential field $\mathbf{A}$ one will obtain for a given driving currents configuration might differ from the ideal case considered above. Take for example $\mathbf{A_s}$. Because the inductive electric field $-\partial \mathbf{A_s}/\partial t$ is along the static magnetic field lines, screening is expected to take place, and the effective $\mathbf{A_s}$ will then be a function of position~\cite{Rax2016}. The actual vector potential will be even more complex if collisional effects are to be accounted for.

\section{Summary}
\label{Sec:IV}

Innovative separation technologies could offer ingenious solutions to important societal challenges. One example of innovative separation technology is plasma separation. Plasma separation stands out from conventional separation techniques by allowing separation at the elemental level based on physical properties. Once a plasma is made out of the mixture to be separated, all differential transport and confinement properties found in plasmas can be leveraged to produce separation. In essence, plasma separation is an extension of plasma confinement physics, but for which the focus is shifted from maximizing confinement to maximizing differential effects.

One particular physical criteria for separation in a plasma is atomic mass. Plasma mass separation could prove valuable for nuclear waste cleanup, nuclear spent fuel reprocessing and rare earth recycling. Although isotope separation motivated the development of a few plasma mass filtration concepts in the 1980s, most of these concepts feature limited throughput. Most often, this limit results from constrains on plasma density set by collisions. Since new applications require processing large quantities of material, there is a need for developing new plasma mass separation concepts.

Although mass separation can be envisioned in many ways, rotating plasmas hold promise owing to centrifugal forces. Rotating plasmas can be used similarly to spinning gases or liquids to separate elements in plasma centrifuges. However, the uniqueness of plasmas lies in the fact that other forces can be leveraged in combination with centrifugal forces. This obviously includes electric and magnetic forces, but also mirror forces. A particle in a spinning gas column sees a mass dependent parabolic centrifugal potential. On the other hand, a charged particle in an $\mathbf{E}\times\mathbf{B}$ spinning plasma column (axial magnetic field, radial electric field) sees the same parabolic centrifugal potential plus an electric potential which depends on the applied potential radial profile. The extra control knob offered by the applied electric potential allows to confine radially light ions while deconfining heavy ions, or to separate light and heavy ions in distinct radial potential wells. Further means of control on particle dynamics can be obtained if abandoning the purely axial magnetic field topology. One solution then consists in combining magnetic pressure with variation of centrifugal potential along the magnetic field line to create mass differential confinement properties. Finally, another way to produce plasma rotation consists in using a rotating magnetic field. This can in principle be achieved through different driving currents configurations, which each leads to different mass confinement capabilities.

The development of actual plasma mass filtering devices hinges on the demonstration of the ability to produce and control transverse electric fields for $\mathbf{E}\times\mathbf{B}$ rotating devices, or suitable potential vector fields for rotating magnetic field configurations. Most importantly, this capability will have to be demonstrated for plasma parameters which are compatible with high-throughput separation applications. To the extent that perpendicular transport and rotation play a key role on performances in a large number of cross-field devices, it is anticipated that these results will benefit many applications besides mass separation.

\section*{Acknowledgments}
The authors would like to thank Dr. F. Levinton, Dr. M. Galante and I. E. Ochs for constructive discussions.

\section*{References}
\providecommand{\newblock}{}

\end{document}